# Revolutionising inverse design of magnesium alloys through generative adversarial networks

Marzie Ghorbani[1,*], Zhipeng Li[2], Nick Birbilis[3,*]

[1]Department of Materials Science and Engineering, Monash University, VIC. 3800, Australia.
[2]College of Engineering, Computing and Cybernetics, The Australian National University, ACT. 2601, Australia.
[3]Faculty of Engineering, Science and the Built Environment, Deakin University, VIC. 3125, Australia.

[*]marzie.ghorbani@monash.edu,
[*]nick.birbilis@deakin.edu.au

**Abstract**

The utility of machine learning (ML) techniques in materials science has accelerated materials design and discovery. However, the accuracy of ML models - particularly deep neural networks - heavily relies on the quality and quantity of the training data. Data collection methods often have limitations arising from cost, difficulty, and resource-intensive human efforts. Thus, limited high-quality data, especially for novel materials, poses a significant challenge in developing reliable ML models. Generative adversarial networks (GANs) offer one solution to augment datasets through synthetic sample generation. The present work explores the application of GANs in magnesium (Mg) alloy design, by training two deep neural networks within the structure of a GAN to generate new (novel) alloys with desired mechanical properties. This data augmentation-based strategy contributes to model robustness, particularly in cases where traditional data collection is impractical. The approach presented may expedite Mg alloy development, through a GAN assisted inverse design approach.

**Keywords:** Magnesium alloys, inverse design, generative adversarial networks, dataset augmentation

## 1. Introduction

The past few years have corresponded to a significant upsurge in the pursuit of accelerating materials design and discovery through the integration of machine learning (ML) techniques [1-5]. This convergence has given rise to an emerging interdisciplinary research domain known as materials informatics. Within this field, the synergistic blend of materials science and data-driven methodologies has paved the way for an upswing in the identification of novel materials. This evolving landscape highlights the potential of ML approaches to revolutionise materials discovery processes and achieve impactful advancements [6, 7]. The materials informatics approach is unique to computational materials science approaches, the latter which seek to determine materials priorities using techniques such as first principles calculations, or mechanistic models [8]. In the case of complex materials which includes engineering alloys, one can argue that mechanistic details are embedded within the associated alloy property data. Such property data, related through complex compositions and processing routes, is too complex for human interpretation *en masse*, but complex underlying relations may be appropriately interpreted via ML algorithms. Naturally, there is a scope for both materials informatics and mechanistic model to be integrated – leveraging these approaches; particularly as computation speeds continue to increase in the future.

The availability of data is critical to materials design and discovery through ML. While striving to maintain data quality, the magnitude of the training dataset plays a pivotal role in elevating the precision of trained ML models, notably artificial neural networks (ANNs) [9, 10]. Nonetheless, a recurring challenge arises across various materials investigations, particularly those focused on novel materials, wherein the scarcity of high-quality and high-volume data persists, impeding the development of reliable ML models [6, 11, 12]. In addition, the acquisition of experimental data, referred to "real" or empirical data for building extensive and high-quality training datasets has substantial hurdles, primarily due to the difficulty, high cost, and the considerable human effort and expertise required for its collection [6, 13, 14].

In response to the issue of limited or inadequate training data availability, the utility of generative adversarial networks (GANs) has recently gained noteworthy traction as a viable solution for data augmentation [15, 16]. Generative networks, specifically GANs, were first introduced by Goodfellow in 2014 [17], however their practical use has gained considerable momentum since ~2017, where GANs were employed to generate realistic synthetic images [18, 19] and enhance the quality of existing images [20]. Within the GAN framework, two neural networks coexist, namely a *generator* and a *discriminator* network. These networks engage computationally in a dynamic interplay as they learn from the real data distribution and partake in what is described as a competitive process against one other [21]. The result of this interplay is that a proficiently executed and hence trained GAN that has the capability to produce synthetic data which aligns with the same distribution as the original (real) training dataset - a technique commonly referred to as a data augmentation-based strategy [15, 22]. This strategy holds the potential to enhance the robustness and efficacy of models in scenarios where a problem is complex and traditional data collection methods prove either impractical or costly [6, 23] such as in biology [24, 25], structural engineering [26, 27] and material science [28-32]. An illustration of the structure of a GAN framework is presented in Figure 1.

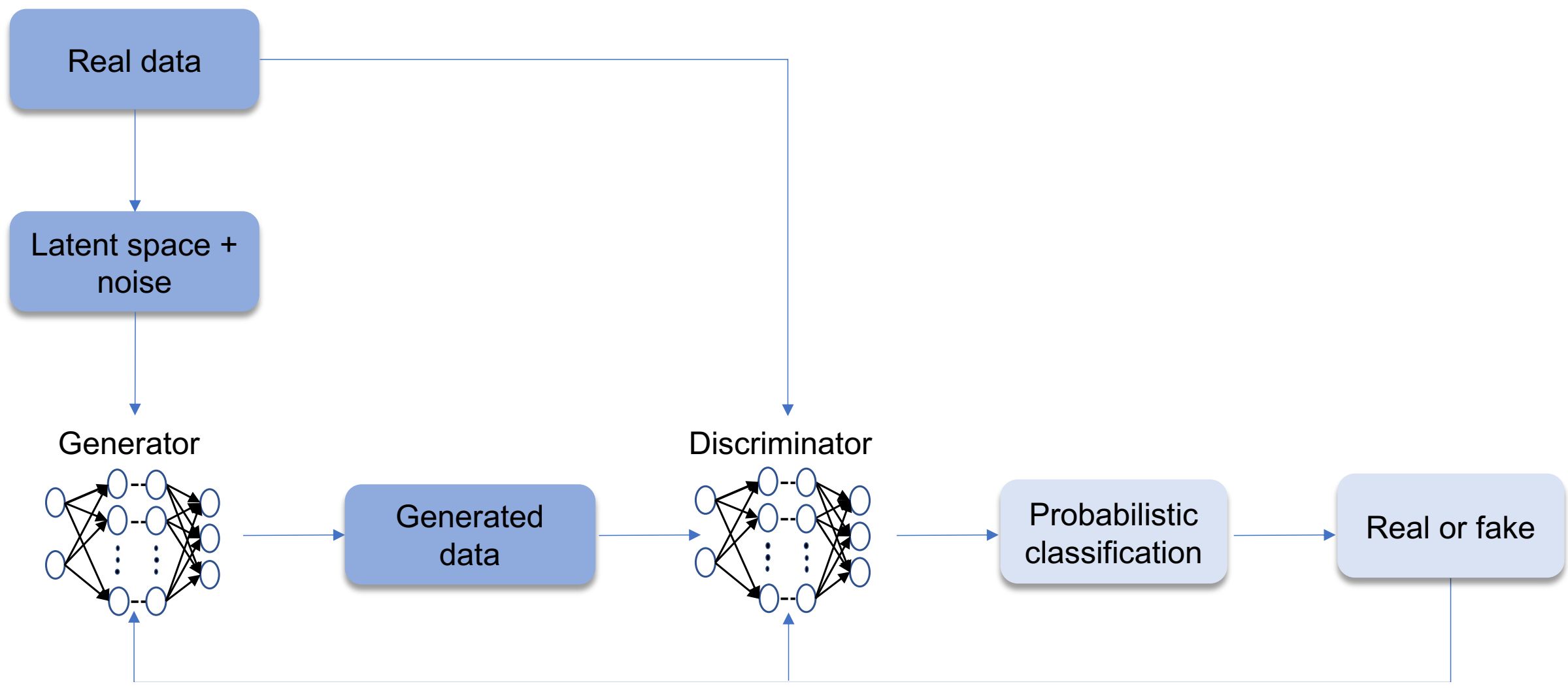

*Figure 1: Illustration of the structure of a generative adversarial network, encompassing generator and discriminator neural networks.*

In the field of materials science, deep learning-based generative models have found practical application in inverse materials design, enabling the discovery of new materials with specifically desired functionalities [1, 23, 33, 34]. In fact, inverse materials design entails identifying the optimal material for a specific application [2, 7, 35] – where the desired properties are known, and the determination of a fit-for-purpose material is the desired model ‘output’. Relevant works in this domain commenced with a focus on inorganic materials and their crystal structure optimisation [33, 36], and more recently expanded to molecular discoveries [37], rupture life and creep rate prediction of high temperature alloys [23], multi principal element alloys and their desired phase prediction [12], multiaxial fatigue life prediction for 316L stainless steel [38], magnetic field prediction of the magnetic materials [39], phase separation structure design in polymer alloys [40], hardness prediction of high entropy alloy [29], aging phenomena and deterioration modelling [41], deep drawing process simulation [42], microstructural evolution of a nickel-alloy (A230) [43], concrete compressive strength prediction [44], crack detection in Li-ion electrodes [45], resistance spot welding quality inspection [46], and high-resolution surface morphology image generation of titanium alloys [47]. In a recent study, generative networks were employed to determine the relationship between tensile properties and hole expansion rate in highly formable materials that include steels and copper alloys [48]. Hence, in summary, while GANs have made significant strides in materials science, particularly in polymers, photonics materials, inorganic solid-state functional materials, porous materials, and thermoelectric materials [49] [50].It is noted that, at present, the application of GANs in metallic alloy design with desired mechanical properties is still relatively limited [51]. The predominant focus of ML research on metallic alloys lies in the materials-to-property prediction route [49, 52] rather than optimisation and inverse design [1, 53].

In this work, by training a GAN on a magnesium (Mg) alloy dataset curated by the authors [54], it is anticipated that the network may learn the underlying relationships and patterns in the Mg-alloy dataset, allowing to the model to generate novel alloy compositions with desired properties. Interest in Mg-alloys remains significant, owing to Mg-alloys being the lightest of the engineering materials known as structural alloys [55]. The present study seeks to explore the application of GANs for inverse design of Mg alloys. The use of GANs may significantly reduce the time and cost involved in the alloy development process by facilitating generation of new compositions with desired properties – which may then be evaluated using computational simulations and experimental validation (including human reasoning) to assess their performance [56]. The workflow in this study, which is shared via data availability of the methods used, holds potential to significantly impact the field of alloy design [57].

## 2. Methods

### *2.1. Inverse prediction via generative adversarial networks (GANs)*

Generative adversarial networks consist of a generator ($G$) and a discriminator ($D$) neural network [58, 59]. These networks/models operate through a simultaneous learning process of the complex high-dimensional probability distribution of input data [60], with one model (the generator) focused on generating a random search space and the other model (the discriminator) committed to differentiate between generated samples and real data [34, 61, 62]. In the architecture of a GAN, the generator plays a critical role. The generator takes a stochastic noise vector $z$ as input. The generators task is to manipulate this input noise vector, denoted as $G(z)$, to synthesise novel samples. The synthesised samples (i.e., data points) then traverse to the discriminator. The discriminator, serving as a discerning binary classifier, undertakes the task of differentiating between real and synthetic samples, collectively evaluating both real and synthetic instances produced by the generator. By analysing the outcomes, the discriminator contributes to the dynamic update of the generator's parameters [63, 64]. This dynamic, characterised by an ongoing quest to surpass each other's performance, forms the core of the adversarial aspect in GANs. This so-called adversarial nature gives rise to a minimax game, wherein each entity strives to outperform the other, serving as the basis for the optimisation process [1]. Throughout the iterative training process, the parameters of both the generator and the discriminator undergo refinement through the principle of backpropagation. The overarching objective is to empower the generator to produce realistic data points, while concurrently sharpening the proficiency of the discriminator in distinguishing real data from synthetic data. Specifically, this drives the generator to minimise a specific loss function, while the discriminator seeks to maximise it - a concept encapsulated by the Minimax loss [37, 65]. The convergence of these dynamics results in the generator producing progressively more realistic data and becoming increasingly proficient at discerning real data from synthetic data - the competition proceeding until the synthetic data becomes indistinguishable from real data. New (novel) alloy compositions may be generated via an adversarial process, whereby the Minimax loss is given as:

$$Min_G \, Max_D \, f(D, G) = E_{x \sim P_{data}} \left[ log\big(D(x)\big) \right] + E_{x \sim P_z} \left[ log\Big(1 - D\big(G(z)\big)\Big) \right] \quad \text{(Eqn. 1)}$$

where $D(x)$ is the probability estimate of the input sample data $x$ (from real data), while $G(z)$ is the output of the generator for a given random noise vector $z$ and the discriminators probability estimate if the synthetically generated data is real, is shown as $D\big(G(z)\big)$. The term $E_{x\sim P_{data}}$ is the expected value over all real data samples, and $E_{x\sim P_z}$ is the expected value over all random inputs to the generator; whilst $P_z$ is a noise distribution [15, 65, 66]. Novel samples may be generated on a non-biased scheme, as GANs rely on the discrepancies between synthetic and real data points instead of the likelihood of the data distribution [35]. Herein, a GAN model was employed to generate new alloy compositions by learning the data distribution from a (real) Mg alloy dataset.

### *2.2. Dataset*

An open access Mg alloy dataset including 916 datapoints was used to train the developed GAN model. The input features encompassed chemical composition and thermomechanical routs of the alloys, while the ultimate and yield strengths and ductility were included as target mechanical properties. This dataset of high dimensionality contains 31 alloying elements and 6 processing categories and is available from [54]. Two deep generative networks of **G** and **D** within the structure of GAN model, were trained with this Mg-alloy data.

### *2.3. Prediction performance of feed-forward ML model*

To assess the predictive performance of feed-forward ML models with GAN-augmented Mg alloy dataset, four regression models with different levels of complexity were employed. The models were trained with the original and GAN-augmented Mg-alloy datasets, then predictive performance was computed for mechanical properties of UTS and ductility. The four models explored were:

i) Least absolute shrinkage and selection operator with information criterion regularization (LassoIC),
ii) Kernel ridge regression (KRR),
iii) Random forest (RF),
iv) Neural network (NN).

The dataset was divided into training and testing sets using a 9:1 ratio, facilitating the assessment of model performance on previously unseen data. This approach allows an evaluation of the predictive capacity of any model developed, on data that were not part of model training – with the aim of simulating their 'real-world' performance. It is noted that achieving satisfactory model performance solely based upon the training data does not assure a model's predictive ability in practical applications.

The performance metric of the coefficient of determination ($R^2$) was averaged over all iterations.

$$R^2 = 1 - \frac{\sum_i (y_i - \hat{y}_i)^2}{\sum_i (y_i - \bar{y})^2} \qquad \text{Eqn. (2)}$$

where $y_i$ is the value of the target property for the $i^{th}$ data point, $\hat{y}_i$ is the model-predicted value of the target property for the $i^{th}$ data point, and $\bar{y}$ is the mean of the target property over all data points considered [67].

To provide a more robust estimate of model performance, the predictive performance was computed herein through a grid-search cross-validation (GSCV) [68] estimator, by evaluating the model on multiple partitions of the data within 10 repetitions (k = 10). The performance metric was then averaged over all 10 iterations.

## 3. Results and Discussion

### *3.1 Results*

The application of the abovementioned GAN workflow was deployed, and Figure 2 shows the properties of newly predicted novel Mg alloys (depicted by orange dots) that were generated by the GAN within five experiments versus the properties of real existing Mg alloys (blue dots). The properties of interest in the example GAN application depicted by Figure 2 are the Mg alloy ultimate tensile strength (UTS) and the alloy ductility (in %).

As it is evident, the developed GAN for Mg alloys is capable of augmenting the real dataset, which is important due to the rather small size of the existing Mg dataset – in spite of the dataset being the most comprehensive treatise available for Mg-alloys. As is widely appreciate in the field, ML model training is often data greedy upon the requirement of real data; and this data augmentation approach may be a useful ongoing means to improve Mg-alloy design from inverse design relying upon meeting target properties. The concise nature of reporting of the present work does not provide all the model code within the manuscript, and instead readers are directed to the data availability statement which directs them towards a repository for the GAN utilised in this work (which is open access).

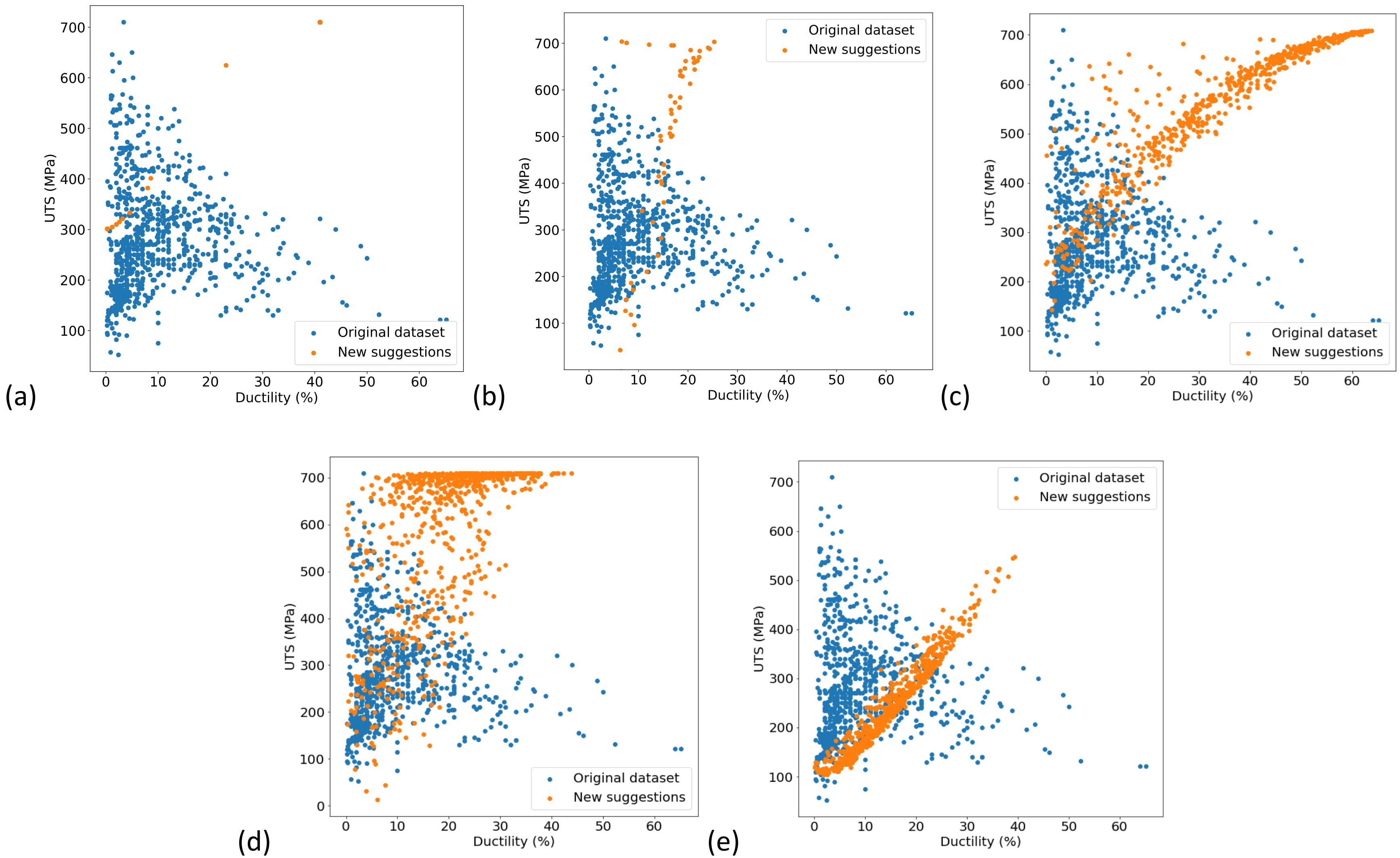

*Figure 2: Proposed new Mg alloys (orange dots) and existing alloys (blue dots) are plotted in terms of UTS versus ductility as generated by the GAN model within five experiments represented by (a) to (e).*

It is noted that from Figure 2, that data generation by employing a GAN approach can be faster than the other conventional ML algorithms, high-throughput virtual screening, and global optimisation techniques [5]. As demonstrated herein, novel generated samples can be used for inverse design of Mg alloys by pre-defining the target mechanical properties and hence narrowing the alloy search space and assisting in accelerated new alloys discovery.

An abridged presentation of GAN derived novel Mg-alloys from the data depicted in figure 2, is provided in Table 1. The table (Table 1) summarises the chemical composition and thermomechanical processing category of proposed novel Mg alloys designed solely by the GAN framework, and of which are believed to have UTS values above 400 MPa and ductility values above 25%.

*Table 1: Chemical composition and thermomechanical category of the proposed new Mg alloys generated by GAN through inverse design. Mechanical properties of >400 MPa for UTS and >25% for ductility are presented.*

| Mg | Nd | Ce | La | Zn | Sn | Al | Ca | Zr | Ag | Ho | Mn | Y | Gd | Cu | Si | Li | Yb | Th | Sb | Pr | Ga | Be | Fe | Ni | Sc | Tb | Dy | Er | Sr | Bi | UTS | Ductility | Processing |
|---|---|---|---|---|---|---|---|---|---|---|---|---|---|---|---|---|---|---|---|---|---|---|---|---|---|---|---|---|---|---|---|---|---|
| 74.7 | 0.6 | 0.8 | 3.5 | 10.1 | 5.5 | 2.8 | 0.0 | 0.0 | 0.0 | 0.1 | 0.0 | 0.0 | 0.0 | 1.3 | 0.0 | 0.0 | 0.0 | 0.0 | 0.0 | 0.0 | 0.0 | 0.0 | 0.0 | 0.0 | 0.0 | 0.0 | 0.0 | 0.0 | 0.0 | 0.7 | 660 | 46 | Extruded |
| 76.5 | 0.0 | 0.0 | 0.0 | 0.0 | 0.0 | 0.0 | 0.0 | 0.0 | 0.0 | 0.0 | 0.0 | 0.0 | 0.0 | 0.7 | 0.0 | 0.0 | 0.1 | 1.0 | 0.1 | 0.0 | 0.0 | 0.0 | 0.0 | 2.6 | 0.0 | 3.5 | 9.0 | 4.7 | 1.7 | 0.0 | 541 | 35 | Extruded |
| 70.3 | 5.7 | 2.8 | 4.2 | 10.1 | 5.7 | 0.0 | 0.0 | 0.0 | 0.0 | 0.0 | 0.0 | 0.0 | 0.0 | 0.2 | 0.0 | 0.0 | 0.0 | 0.0 | 0.0 | 0.4 | 0.0 | 0.0 | 0.0 | 0.0 | 0.0 | 0.0 | 0.0 | 0.0 | 0.0 | 0.6 | 656 | 44 | Extruded |
| 81.5 | 0.0 | 0.0 | 0.0 | 8.5 | 5.5 | 0.0 | 0.0 | 0.1 | 0.0 | 0.0 | 0.0 | 0.0 | 0.0 | 0.0 | 0.0 | 0.0 | 0.0 | 0.0 | 0.0 | 0.2 | 0.0 | 0.0 | 0.0 | 0.0 | 0.0 | 0.0 | 0.0 | 3.0 | 0.5 | 0.7 | 675 | 51 | Extruded |
| 84.4 | 0.0 | 0.0 | 0.0 | 0.0 | 0.0 | 0.0 | 0.0 | 0.0 | 0.0 | 0.0 | 0.0 | 0.0 | 0.0 | 0.0 | 0.0 | 2.3 | 0.0 | 0.0 | 0.0 | 0.2 | 0.0 | 0.0 | 0.0 | 2.6 | 0.0 | 0.0 | 4.9 | 4.5 | 1.0 | 0.0 | 602 | 40 | Extruded |
| 92.4 | 0.0 | 0.0 | 0.0 | 0.0 | 0.2 | 4.9 | 0.0 | 0.0 | 0.6 | 1.0 | 0.0 | 0.0 | 0.0 | 0.8 | 0.0 | 0.0 | 0.0 | 0.0 | 0.0 | 0.0 | 0.0 | 0.0 | 0.0 | 0.0 | 0.0 | 0.0 | 0.0 | 0.0 | 0.0 | 0.2 | 633 | 41 | Extruded |
| 83.0 | 0.0 | 0.0 | 0.0 | 0.0 | 0.0 | 0.0 | 0.0 | 0.0 | 0.0 | 0.3 | 0.0 | 0.0 | 0.5 | 1.9 | 2.5 | 0.0 | 0.2 | 1.0 | 0.3 | 0.0 | 0.0 | 0.0 | 0.0 | 0.0 | 0.0 | 0.0 | 4.3 | 4.8 | 1.1 | 0.1 | 656 | 49 | Extruded |
| 72.9 | 5.9 | 2.9 | 4.4 | 10.3 | 3.5 | 0.0 | 0.0 | 0.0 | 0.0 | 0.0 | 0.0 | 0.0 | 0.0 | 0.0 | 0.0 | 0.0 | 0.0 | 0.0 | 0.0 | 0.0 | 0.0 | 0.0 | 0.0 | 0.0 | 0.0 | 0.0 | 0.0 | 0.0 | 0.0 | 0.2 | 613 | 36 | Extruded |
| 72.7 | 0.0 | 0.0 | 0.0 | 3.3 | 2.2 | 0.0 | 0.0 | 0.0 | 0.0 | 0.0 | 0.0 | 0.0 | 0.0 | 0.0 | 6.6 | 9.9 | 1.6 | 0.1 | 0.2 | 1.0 | 0.0 | 0.0 | 0.0 | 1.8 | 0.4 | 0.0 | 0.0 | 0.3 | 0.0 | 0.0 | 641 | 45 | Extruded |
| 78.5 | 0.0 | 0.0 | 0.0 | 0.0 | 0.0 | 9.5 | 1.4 | 0.0 | 4.0 | 1.2 | 0.6 | 0.0 | 0.0 | 0.0 | 0.0 | 1.8 | 0.5 | 0.1 | 0.0 | 0.0 | 0.0 | 0.0 | 0.0 | 0.0 | 0.0 | 1.1 | 0.0 | 0.9 | 0.0 | 0.5 | 679 | 52 | Extruded |
| 89.2 | 0.0 | 0.0 | 0.0 | 3.3 | 2.9 | 2.4 | 0.0 | 0.0 | 0.0 | 0.2 | 0.0 | 0.0 | 0.0 | 0.0 | 0.0 | 0.0 | 0.0 | 0.0 | 0.1 | 0.6 | 0.0 | 0.0 | 0.0 | 0.0 | 0.0 | 0.0 | 0.0 | 0.5 | 0.0 | 0.9 | 682 | 52 | Extruded |
| 76.3 | 0.0 | 0.0 | 0.0 | 2.4 | 4.0 | 10.4 | 1.5 | 0.0 | 2.6 | 0.9 | 0.4 | 0.0 | 0.0 | 0.0 | 0.0 | 0.0 | 0.0 | 0.0 | 0.0 | 0.0 | 0.0 | 0.0 | 0.0 | 0.0 | 0.0 | 0.0 | 0.0 | 0.0 | 0.1 | 1.4 | 693 | 55 | Extruded |
| 84.0 | 0.0 | 0.0 | 0.0 | 1.4 | 1.9 | 0.0 | 0.0 | 0.0 | 0.0 | 0.0 | 0.0 | 0.0 | 1.2 | 1.0 | 0.0 | 0.0 | 0.0 | 0.0 | 0.0 | 0.0 | 0.0 | 0.0 | 0.0 | 0.0 | 0.0 | 0.0 | 4.5 | 4.7 | 1.2 | 0.1 | 643 | 46 | Extruded |
| 91.4 | 0.0 | 0.0 | 0.0 | 0.0 | 0.0 | 2.8 | 0.0 | 0.0 | 0.0 | 0.0 | 0.0 | 0.0 | 0.0 | 0.0 | 0.0 | 0.0 | 0.0 | 0.0 | 0.0 | 0.0 | 0.0 | 0.0 | 0.0 | 0.0 | 0.0 | 0.0 | 1.1 | 4.0 | 0.6 | 0.2 | 644 | 46 | Extruded |
| 81.2 | 0.0 | 0.0 | 0.0 | 0.0 | 0.0 | 1.1 | 0.0 | 0.0 | 0.0 | 0.7 | 0.0 | 0.0 | 0.0 | 1.0 | 2.6 | 2.6 | 1.6 | 2.2 | 0.4 | 0.0 | 0.0 | 0.0 | 0.0 | 0.0 | 0.0 | 0.0 | 2.2 | 4.1 | 0.3 | 0.0 | 533 | 32 | Extruded |
| 79.9 | 0.0 | 0.0 | 0.0 | 0.0 | 0.0 | 7.8 | 0.0 | 0.0 | 4.5 | 1.4 | 0.8 | 0.0 | 0.0 | 0.8 | 0.4 | 0.0 | 1.6 | 2.5 | 0.5 | 0.0 | 0.0 | 0.0 | 0.0 | 0.0 | 0.0 | 0.0 | 0.0 | 0.0 | 0.0 | 0.0 | 598 | 37 | Extruded |
| 87.9 | 0.0 | 0.0 | 0.0 | 0.0 | 0.0 | 7.7 | 0.0 | 0.0 | 2.8 | 1.2 | 0.4 | 0.0 | 0.0 | 0.0 | 0.0 | 0.0 | 0.0 | 0.0 | 0.0 | 0.0 | 0.0 | 0.0 | 0.0 | 0.0 | 0.0 | 0.0 | 0.0 | 0.0 | 0.0 | 0.0 | 560 | 30 | Extruded |
| 75.2 | 0.0 | 0.0 | 2.4 | 9.7 | 6.2 | 3.8 | 0.0 | 0.0 | 0.0 | 0.0 | 0.0 | 0.0 | 0.0 | 0.0 | 0.0 | 0.0 | 0.0 | 0.0 | 0.0 | 0.3 | 0.0 | 0.0 | 0.0 | 0.0 | 0.2 | 1.4 | 0.0 | 0.0 | 0.0 | 0.9 | 671 | 48 | Extruded |
| 86.6 | 0.0 | 0.0 | 0.0 | 0.0 | 2.2 | 7.8 | 0.0 | 0.0 | 0.5 | 0.6 | 0.0 | 0.0 | 0.0 | 0.0 | 0.0 | 0.0 | 0.0 | 0.0 | 0.0 | 0.1 | 0.0 | 0.0 | 0.0 | 0.0 | 0.0 | 0.0 | 0.0 | 0.9 | 0.1 | 1.3 | 694 | 56 | Extruded |
| 73.4 | 0.0 | 0.0 | 0.0 | 0.0 | 0.0 | 0.0 | 0.0 | 0.0 | 0.0 | 0.0 | 0.0 | 0.0 | 3.4 | 1.0 | 0.0 | 0.0 | 0.0 | 0.0 | 0.0 | 0.0 | 0.0 | 0.0 | 0.0 | 3.1 | 0.1 | 4.1 | 8.7 | 4.5 | 1.7 | 0.0 | 505 | 32 | Extruded |
| 75.8 | 0.0 | 0.0 | 0.0 | 0.0 | 0.0 | 0.0 | 0.0 | 0.0 | 0.0 | 0.0 | 0.0 | 0.6 | 0.0 | 0.0 | 1.9 | 5.9 | 0.5 | 0.0 | 0.0 | 0.5 | 0.0 | 0.0 | 0.0 | 2.8 | 0.0 | 0.0 | 6.4 | 4.5 | 1.1 | 0.0 | 517 | 31 | Extruded |
| 70.9 | 0.3 | 0.6 | 3.1 | 9.3 | 5.3 | 6.4 | 0.0 | 0.0 | 1.3 | 0.9 | 0.0 | 0.0 | 0.0 | 0.8 | 0.0 | 0.0 | 0.0 | 0.0 | 0.0 | 0.0 | 0.0 | 0.0 | 0.0 | 0.0 | 0.0 | 0.0 | 0.0 | 0.0 | 0.0 | 1.2 | 686 | 53 | Extruded |
| 88.5 | 0.0 | 0.0 | 0.0 | 1.0 | 3.0 | 5.7 | 0.0 | 0.0 | 0.0 | 0.2 | 0.0 | 0.0 | 0.0 | 0.0 | 0.0 | 0.0 | 0.0 | 0.0 | 0.0 | 0.0 | 0.0 | 0.0 | 0.0 | 0.0 | 0.0 | 0.0 | 0.0 | 0.5 | 0.0 | 1.0 | 683 | 52 | Extruded |
| 75.1 | 0.0 | 0.0 | 0.0 | 0.0 | 2.1 | 0.2 | 0.0 | 0.0 | 0.0 | 0.0 | 0.0 | 0.0 | 0.0 | 0.0 | 0.0 | 4.9 | 0.1 | 0.0 | 0.2 | 0.9 | 0.5 | 0.0 | 0.0 | 3.5 | 0.0 | 0.0 | 5.8 | 4.3 | 1.4 | 1.0 | 695 | 58 | Extruded |
| 74.1 | 0.0 | 0.0 | 1.8 | 8.8 | 5.6 | 2.7 | 0.0 | 0.0 | 0.0 | 0.0 | 0.0 | 0.0 | 0.0 | 0.0 | 0.0 | 2.7 | 0.0 | 0.0 | 0.3 | 0.9 | 0.5 | 0.0 | 0.0 | 0.0 | 0.0 | 0.0 | 0.0 | 1.3 | 0.2 | 1.3 | 701 | 59 | Extruded |
| 83.9 | 0.0 | 0.0 | 0.0 | 0.0 | 0.0 | 6.7 | 0.0 | 0.0 | 3.1 | 1.1 | 0.5 | 0.0 | 0.0 | 0.0 | 0.0 | 0.0 | 0.6 | 1.4 | 0.1 | 0.0 | 0.0 | 0.0 | 0.0 | 0.0 | 0.0 | 0.0 | 0.0 | 2.6 | 0.0 | 0.0 | 594 | 38 | Extruded |
| 74.5 | 0.7 | 0.6 | 4.1 | 11.2 | 6.1 | 0.0 | 0.0 | 0.0 | 0.0 | 0.0 | 0.0 | 0.0 | 0.0 | 0.0 | 0.0 | 1.8 | 0.0 | 0.0 | 0.1 | 0.7 | 0.0 | 0.0 | 0.0 | 0.0 | 0.0 | 0.0 | 0.0 | 0.0 | 0.0 | 0.2 | 656 | 46 | Extruded |
| 82.5 | 0.0 | 0.0 | 0.0 | 0.0 | 0.0 | 1.7 | 0.0 | 0.0 | 0.0 | 0.0 | 0.0 | 0.0 | 0.0 | 0.0 | 0.0 | 0.0 | 0.0 | 0.0 | 0.0 | 0.1 | 0.0 | 0.0 | 0.0 | 1.1 | 0.0 | 0.0 | 7.9 | 4.9 | 1.6 | 0.4 | 671 | 52 | Extruded |
| 76.1 | 0.0 | 0.0 | 0.0 | 0.0 | 0.0 | 8.9 | 0.8 | 0.0 | 3.0 | 1.1 | 0.3 | 0.0 | 0.0 | 0.0 | 0.0 | 0.0 | 0.0 | 0.0 | 0.0 | 0.0 | 0.0 | 0.0 | 0.0 | 0.0 | 1.7 | 7.5 | 0.2 | 0.0 | 0.0 | 0.5 | 662 | 47 | Extruded |
| 84.6 | 0.0 | 0.0 | 0.0 | 0.0 | 0.0 | 0.0 | 0.0 | 0.0 | 0.0 | 0.0 | 0.0 | 0.0 | 0.0 | 0.0 | 4.0 | 9.1 | 1.6 | 0.3 | 0.0 | 0.4 | 0.0 | 0.0 | 0.0 | 0.0 | 0.0 | 0.0 | 0.0 | 0.0 | 0.0 | 0.0 | 610 | 39 | Extruded |
| 75.9 | 0.0 | 0.0 | 0.0 | 0.0 | 0.0 | 1.5 | 0.0 | 0.0 | 0.0 | 0.5 | 0.0 | 0.0 | 0.0 | 0.0 | 0.0 | 1.4 | 0.0 | 0.0 | 0.2 | 0.5 | 0.2 | 0.0 | 0.0 | 1.8 | 0.0 | 2.1 | 8.7 | 4.6 | 1.8 | 0.8 | 696 | 59 | Extruded |
| 71.9 | 0.0 | 0.0 | 0.0 | 0.0 | 0.5 | 12.2 | 6.3 | 1.0 | 4.9 | 1.2 | 1.3 | 0.0 | 0.0 | 0.0 | 0.0 | 0.0 | 0.0 | 0.2 | 0.0 | 0.0 | 0.0 | 0.0 | 0.0 | 0.0 | 0.0 | 0.0 | 0.0 | 0.1 | 0.0 | 0.6 | 673 | 50 | Extruded |
| 88.4 | 0.0 | 0.0 | 0.0 | 0.0 | 0.1 | 0.0 | 0.0 | 0.0 | 0.0 | 0.0 | 0.0 | 0.0 | 0.0 | 0.0 | 0.0 | 5.2 | 0.0 | 0.0 | 0.0 | 0.6 | 0.0 | 0.0 | 0.0 | 2.4 | 0.0 | 0.0 | 0.0 | 3.2 | 0.2 | 0.0 | 625 | 42 | Extruded |
| 73.0 | 0.0 | 0.0 | 0.0 | 0.0 | 0.5 | 8.9 | 1.9 | 0.0 | 3.4 | 1.0 | 0.5 | 0.0 | 0.0 | 0.0 | 0.0 | 4.9 | 0.2 | 0.0 | 0.2 | 0.8 | 0.3 | 0.0 | 0.0 | 0.0 | 0.0 | 0.0 | 0.0 | 2.4 | 0.5 | 1.5 | 706 | 62 | Extruded |
| 79.5 | 0.0 | 0.0 | 1.4 | 9.0 | 5.2 | 0.0 | 0.0 | 0.0 | 0.0 | 0.0 | 0.0 | 0.0 | 0.0 | 0.0 | 0.0 | 0.1 | 0.0 | 0.0 | 0.1 | 0.7 | 0.2 | 0.0 | 0.0 | 0.0 | 0.0 | 0.0 | 0.0 | 2.9 | 0.4 | 0.6 | 674 | 51 | Extruded |
| 89.7 | 0.0 | 0.0 | 0.0 | 0.0 | 0.1 | 2.1 | 0.0 | 0.0 | 0.0 | 0.0 | 0.0 | 0.0 | 0.0 | 0.0 | 0.3 | 6.3 | 0.3 | 0.0 | 0.0 | 0.6 | 0.0 | 0.0 | 0.0 | 0.0 | 0.0 | 0.0 | 0.0 | 0.3 | 0.0 | 0.4 | 668 | 49 | Extruded |
| 76.8 | 0.0 | 0.0 | 0.0 | 0.0 | 0.0 | 3.9 | 0.0 | 0.0 | 0.0 | 0.3 | 0.0 | 0.0 | 0.0 | 0.0 | 0.9 | 5.9 | 0.5 | 0.0 | 0.0 | 0.2 | 0.0 | 0.0 | 0.0 | 0.0 | 4.9 | 6.6 | 0.0 | 0.0 | 0.0 | 0.0 | 640 | 43 | Extruded |
| 80.4 | 0.0 | 0.0 | 0.0 | 0.0 | 0.0 | 8.7 | 0.2 | 0.0 | 3.3 | 1.1 | 0.4 | 0.0 | 0.0 | 0.0 | 0.0 | 0.0 | 0.4 | 0.7 | 0.0 | 0.0 | 0.0 | 0.0 | 0.0 | 0.0 | 0.1 | 4.6 | 0.0 | 0.0 | 0.0 | 0.0 | 578 | 34 | Extruded |
| 94.5 | 0.0 | 0.0 | 0.0 | 0.0 | 0.0 | 5.3 | 0.0 | 0.0 | 0.0 | 0.0 | 0.0 | 0.0 | 0.0 | 0.0 | 0.0 | 0.0 | 0.0 | 0.0 | 0.0 | 0.0 | 0.0 | 0.0 | 0.0 | 0.0 | 0.0 | 0.0 | 0.0 | 0.0 | 0.0 | 0.2 | 644 | 44 | Extruded |
| 76.7 | 0.0 | 0.0 | 0.0 | 0.0 | 0.0 | 8.2 | 2.3 | 0.0 | 2.6 | 0.9 | 0.4 | 0.0 | 0.0 | 0.0 | 0.0 | 0.0 | 0.0 | 0.0 | 0.0 | 0.0 | 0.2 | 0.0 | 0.0 | 0.0 | 0.0 | 0.0 | 4.1 | 3.9 | 0.7 | 0.0 | 589 | 38 | Extruded |
| 89.1 | 0.0 | 0.0 | 0.0 | 2.4 | 3.7 | 3.1 | 0.0 | 0.0 | 0.0 | 0.0 | 0.0 | 0.0 | 0.0 | 0.0 | 0.0 | 0.6 | 0.0 | 0.0 | 0.0 | 0.5 | 0.0 | 0.0 | 0.0 | 0.0 | 0.0 | 0.0 | 0.0 | 0.1 | 0.0 | 0.6 | 668 | 49 | Extruded |
| 79.4 | 0.0 | 0.0 | 0.0 | 0.0 | 0.0 | 0.0 | 0.0 | 0.0 | 0.0 | 0.0 | 0.0 | 0.0 | 0.0 | 0.0 | 4.0 | 8.0 | 1.5 | 0.6 | 0.1 | 0.2 | 0.0 | 0.0 | 0.0 | 0.0 | 0.0 | 0.0 | 2.1 | 3.8 | 0.4 | 0.0 | 581 | 37 | Extruded |
| 84.9 | 0.0 | 0.0 | 0.2 | 7.6 | 5.1 | 0.8 | 0.0 | 0.0 | 0.0 | 0.0 | 0.0 | 0.0 | 0.0 | 0.0 | 0.0 | 0.0 | 0.0 | 0.0 | 0.0 | 0.4 | 0.0 | 0.0 | 0.0 | 0.0 | 0.0 | 0.0 | 0.0 | 0.5 | 0.0 | 0.5 | 668 | 49 | Extruded |
| 71.3 | 0.6 | 0.0 | 0.0 | 1.3 | 0.0 | 4.8 | 0.0 | 0.0 | 0.0 | 0.0 | 0.0 | 0.0 | 0.0 | 0.0 | 0.0 | 0.0 | 0.0 | 0.2 | 0.0 | 0.7 | 0.4 | 0.0 | 0.0 | 0.0 | 0.0 | 12.3 | 0.0 | 4.7 | 1.3 | 2.4 | 708 | 25 | Extruded |
| 81.5 | 0.6 | 0.0 | 0.0 | 2.6 | 0.0 | 5.9 | 0.0 | 0.0 | 0.0 | 0.0 | 0.0 | 0.0 | 0.0 | 0.0 | 0.5 | 0.0 | 0.5 | 0.0 | 0.2 | 0.0 | 0.6 | 0.0 | 0.0 | 0.0 | 0.0 | 0.4 | 0.0 | 4.8 | 0.2 | 2.5 | 686 | 25 | Extruded |
| 74.9 | 0.0 | 0.3 | 0.0 | 7.9 | 0.0 | 11.6 | 0.0 | 0.6 | 0.1 | 0.3 | 0.2 | 0.0 | 0.0 | 0.0 | 0.2 | 0.0 | 0.3 | 0.0 | 0.0 | 0.0 | 0.0 | 0.0 | 0.0 | 0.0 | 0.0 | 0.0 | 0.0 | 2.3 | 0.0 | 1.4 | 650 | 26 | Extruded |
| 73.7 | 2.0 | 0.0 | 0.0 | 0.0 | 0.7 | 0.0 | 1.3 | 0.0 | 1.3 | 0.0 | 0.5 | 0.0 | 0.0 | 0.0 | 0.0 | 0.0 | 0.6 | 0.0 | 0.1 | 0.0 | 0.0 | 0.0 | 0.0 | 1.2 | 0.0 | 9.9 | 0.0 | 5.1 | 1.0 | 2.6 | 702 | 26 | Extruded |

Augmented datasets (from the GAN model) may be used to overcome the sparsity of (real) Mg alloy training data, leading to improved accuracy of predictive ML models. Figures 3 and 4 reveal a comparison of the predictive performance of four ML models trained with the real (denoted as Original) and augmented (denoted by GAN) Mg alloy dataset, for UTS and ductility prediction, respectively.

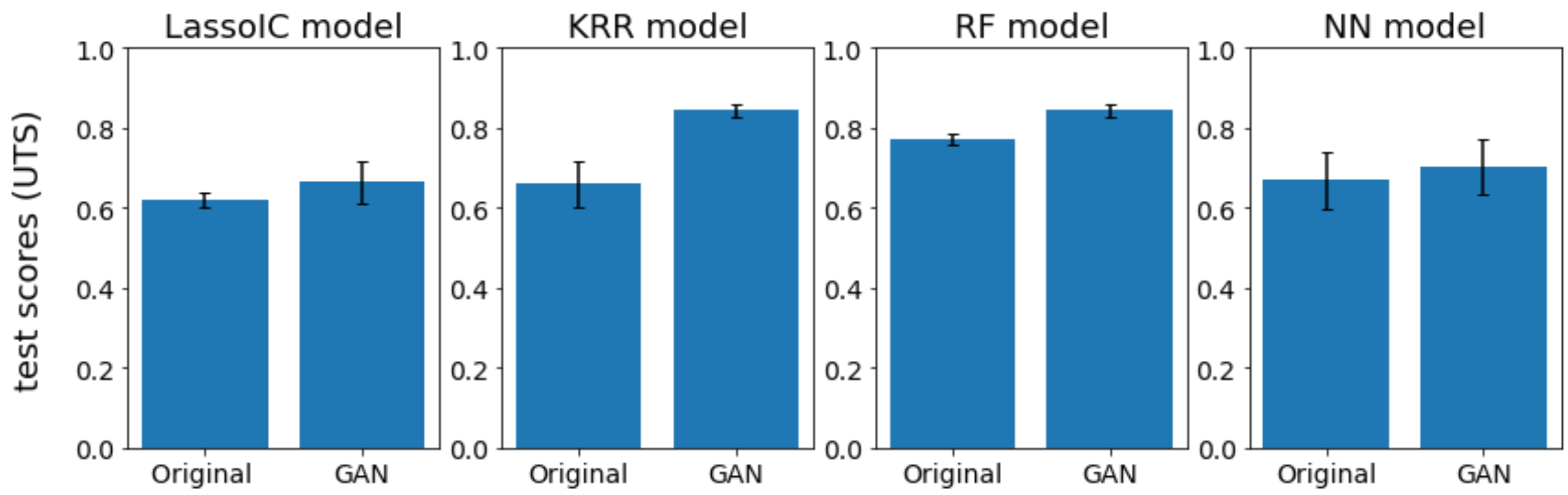

*Figure 3: Comparison of predictive performance of four ML models trained with the real (denoted as Original) and augmented (denoted by GAN) Mg alloy dataset for UTS prediction.*

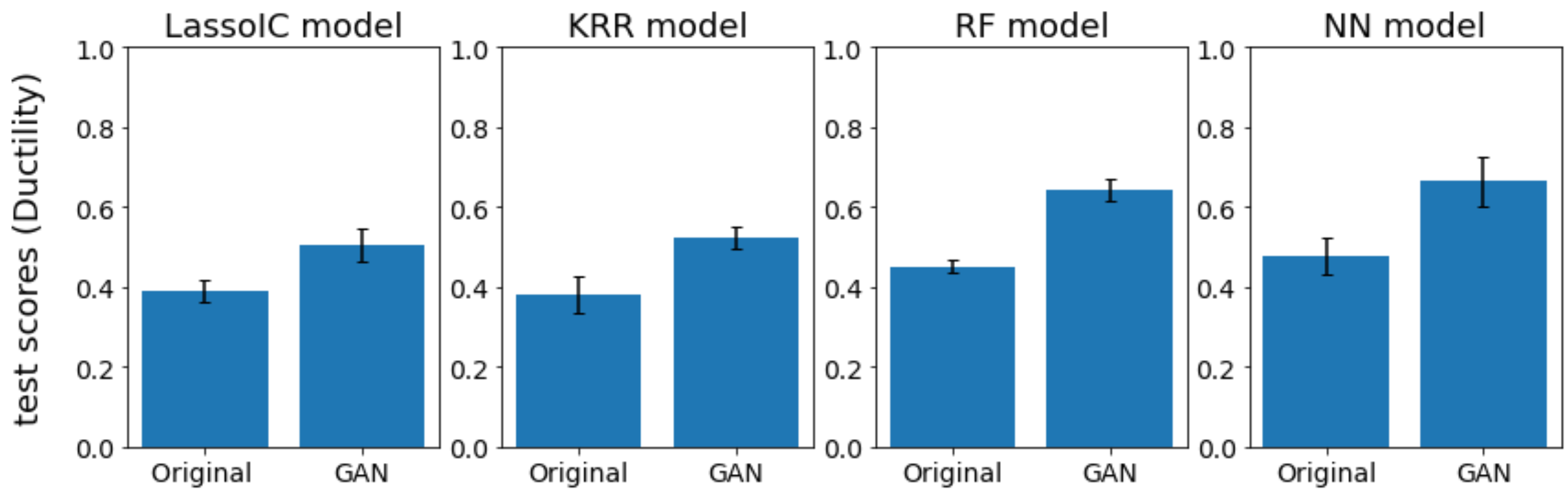

*Figure 4: Comparison of predictive performance of four ML models trained with the real (denoted as Original) and augmented (denoted by GAN) Mg alloy dataset for ductility prediction.*

It is noted that from Figures 3 and 4, hat predictive performance of all four models as increased using the GAN-augmented dataset, for prediction of both the UTS and ductility. In may be posited that GAN augmented data can lead to the improved accuracy of predictive ML models, albeit whilst not alleviating the role of human reasoning in subsequent alloy predictions. The values of the test scores for all the models are summarised in Table 2.

*Table 2: Test score values for UTS and ductility prediction of Mg alloys by four ML models trained with the real (denoted as Original) and augmented (denoted by GAN) Mg alloy dataset.*

| Model | UTS prediction | | Ductility prediction | |
|---|---|---|---|---|
| | Original dataset | GAN-augmented dataset | Original dataset | GAN-augmented dataset |
| LassoIC | 0.62 | 0.66 | 0.39 | 0.50 |
| KRR | 0.66 | 0.84 | 0.38 | 0.52 |
| RF | 0.77 | 0.84 | 0.45 | 0.64 |
| NN | 0.67 | 0.70 | 0.48 | 0.66 |

The alloy compositions and processing route predictions by the GAN are done in a high-volume manner. On this basis, the down-selection of alloys in Table 1 that may warrant further investigation for empirical production should be assessed by human-level reasoning. Specifically, because the production of Mg-alloys is highly influenced by the desire to include or avoid certain alloying elements (due to multi-dimensional aspects that vary from price, alloy density and alloy performance).

### *3.2 General Discussion*

While GANs have shown great potential in facilitating inverse design, it is important to recognise the potential limitations of the approach. The challenges related to computational complexity, model interpretability, and the requirement for extensive computational resources for training and implementation are important considerations.

The utilisation of augmented data to train ML models with improved performance, while logical from a data scarcity perspective, may raise concerns regarding the feasibility of the approach in a practical sense more broadly. Understandably, engineers may seek convincing evidence of the reliability and robustness of models trained with augmented data. To establish the feasibility and credibility of these models, it is important to emphasise that the model learns from the training process, while none of the test data was seen by the model. This clarification is crucial to demonstrate the ML models integrity of generalisation ability. Furthermore, and perhaps most obviously, accompanying experimental validation is essential to build trust in the feasibility of these data-driven approaches within the domain of materials science and engineering (and to learn further about any challenges and opportunities).

Based on the implementation of the approach outlined in this study, it is indeed possible that the model will generate some candidates that are potentially problematic or (within a reasonable population) erroneous. The authors, nor likely any users of ML models related to materials properties, do not claim complete accuracy in the relevance of alloy predictions. Although some generated samples in the examples shown in Figure 2 and Table 1 have simultaneous high UTS (> 650 MPa) and high ductility (> 35%) might seem improbable based on Mg-alloys empirically produced to date, their elemental combinations serve two purposes: firstly, they can inform alloy practitioners of potentially novel alloy combinations that are not appropriately tested to date; and, secondly, the alloy compositions may serve as representatives of a potential Mg alloy class with a defined processing route, offering insights and potential new alloy classes. From the authors assessment of the GAN derived alloys more generally, promising Mg alloy families may include:

i. Mg-Gd-Al-Y-Zn
ii. Mg-Gd-Al-Zn
iii. Mg-Gd-Zn-Y
iv. alloys containing blends of what are elements notionally separated due to the prospect that they will produce complex co-precipitation
v. alloys that have specific alloying from one (or more) rare earth element in combination with commodity (i.e., Al or Zn) alloying additions

Undoubtedly, alloys generated by the GAN approach herein should be evaluated further by using computational simulations or (in particular) experimental validation to assess their practical performance. The purpose of the work herein was to provide a concise treatise on the application of GANs as a timely inverse-design tool for Mg-alloys.

**Data availability**

The developed GAN model for Mg alloy dataset augmenting is publicly available at the following repository: https://github.com/katrina-coder/GAN_Mg_Alloy

## 4. Conclusions

In this work, a new method for generating novel Mg alloys via a generative adversarial network (GAN) model was proposed. This is the first such application of GANs towards the inverse-design of Mg alloys. Based on the work herein, it may be concluded that:

- Generative networks may be beneficial in order to overcome the sparsity of Mg alloy (real) training data through dataset augmentation, leading to improved accuracy of the predictive ML models and accelerating new alloy discovery. The present work did not comprehensively explore the benefits of dataset augmentation and its general feasibility though experiments. Nonetheless, for scenarios related to alloy design where there is (training) data scarcity, dataset augmentation merits further study – including its computational significance.
- Novel Mg alloy compositions and their mechanical properties may be generated through a non-biased process, as GANs do not rely on likelihood of the data distribution - since generative models improve their accuracy via detecting discrepancies between real and synthetic data.
- Data generation may be achieved on a timescale faster than the other high-throughput virtual screening and optimisation processes, such as Bayesian optimisation.
- The ability of GANs in expanding design space facilitates inverse design for Mg alloys by user pre-defining desired target properties.

**Acknowledgements**

We acknowledge the support of a Monash-IITB Academy Scholarship. We also thank the Australian Research Council for support via DP190103592.